\documentclass{cimento}
\usepackage[dvips]{graphicx}

 \title{Nonrelativistic Electron-Electron Thermal Bremsstrahlung Gaunt Factor}

 \author{Naoki~Itoh\from{ins:s}\ETC, Youhei~Kawana\from{ins:s} 
         \atque Satoshi~Nozawa\from{ins:j}}

 \instlist{ \inst{ins:s} Department of Physics, Sophia University, 7-1 Kioi-cho, Chiyoda-ku, Tokyo, 102-8554, Japan
 \inst{ins:j} Josai Junior College for Women, 1-1 Keyakidai, Sakado-shi, Saitama, 350-0290, Japan   }

\PACSes{ \PACSit{95.30.Cq}{Elementary particle processes}
         \PACSit{95.30.Gv}{Radiation mechanisms}
         \PACSit{98.65.Cw}{Galaxy clusters}
         \PACSit{98.80-k}{Cosmology}}

\begin{document}

\maketitle

\begin{abstract}

  With the launch of the {\it Chandra X-Ray Observatory} and {\it XMM-Newton}, X-ray astronomy has entered an era of precision science.  The precision observational data taken by these satellites require precision basic physics data for their analysis.  The present authors have recently published a series of papers in which they have reported the results of the accurate calculations of the electron-ion thermal bremsstrahlung which are particularly suited for the analysis of the high-temperature galaxy clusters.  In this paper we will present accurate analytic fitting formulae for the nonrelativistic electron-electron thermal bremsstrahlung Gaunt factor which will be useful for the analysis of the X-ray observational data taken by the {\it Chandra X-Ray Observatory} and {\it XMM-Newton}.  With the publication of the present paper X-ray astronomers will be provided with an accurate thermal bremsstrahlung Gaunt factor which has an overall accuracy better than 1\% for the temperature range 6.0 $\leq$ log $T$[K] $\leq$ 8.5.

\end{abstract}

\section{Introduction}

   With the launch of the {\it Chandra X-Ray Observatory} and {\it XMM-Newton}, X-ray astronomy has entered an era of precision science.  In particular, these satellites have revolutionized the accuracy of the observational data of the galaxy clusters (Allen et al.~\cite{ref:allen}; Schmidt et al.~\cite{ref:schmidt}).  The precision observational data taken by these satellites require precision basic physics data for their analysis.

  The present authors have recently published a series of papers in which they have reported the results of the accurate calculations of the electron-ion thermal bremsstrahlung which are particularly suited for the analysis of the high-temperature galaxy clusters (Nozawa, Itoh, \& Kohyama~\cite{ref:nozawa}; Itoh et al.~\cite{ref:itoh00}; Itoh et al.~\cite{ref:itoh01}).  Their calculations are based on the method of Itoh and his collaborators (Itoh, Nakagawa, \& Kohyama~\cite{ref:itoh85}; Nakagawa, Kohyama, \& Itoh~\cite{ref:nakagawa}; Itoh, Kojo, \& Nakagawa~\cite{ref:itoh90}; Itoh et al.~\cite{ref:itoh91},~\cite{ref:itoh97}).  In Itoh et al.~\cite{ref:itoh00} and Itoh et al.~\cite{ref:itoh01} the present authors have presented accurate analytic fitting formulae for the electron-ion thermal bremsstrahlung Gaunt factors which have a general accuracy of about 0.1\% for the ranges 1 $\leq Z_{j} \leq$ 28, 6.0 $\leq$ log $T$[K] $\leq$ 8.5, where $Z_{j}$ is the charge of the ion and $T$ is the electron temperature.  These electron-ion thermal bremsstrahlung Gaunt factors will be ideally suited for the analysis of the precision X-ray data taken by the {\it Chandra X-Ray Observatory} and {\it XMM-Newton}.

  It has been known for some time that the electron-electron thermal bremsstrahlung makes a percent-order contribution to the total thermal bremsstrahlung at $T \sim 10^{8}$K (Maxon \& Corman~\cite{ref:maxon67}; Maxon~\cite{ref:maxon72}; Haug~\cite{ref:haug}; Svensson~\cite{ref:svensson}; Dermer~\cite{ref:dermer}).  In particular, Haug~\cite{ref:haug} has carried out a relativistic calculation of the electron-electron thermal bremsstrahlung and has found that the nonrelativistic electron-electron thermal bremsstrahlung results calculated by Maxon \& Corman~\cite{ref:maxon67} and by Maxon~\cite{ref:maxon72} agree with his relativistic results within about 1\% accuracy at $T \sim 1.5 \times 10^{8}$K.  Since the electron-electron thermal bremsstrahlung is a percent-order contribution compared with the electron-ion thermal bremsstrahlung, the results of the nonrelativistic electron-electron  bremsstrahlung will be sufficiently accurate for the analysis of the X-ray observational data for the galaxy clusters.

  In this paper we will present accurate analytic fitting formulae for the nonrelativistic electron-electron thermal bremsstrahlung Gaunt factors for the temperature range $-$4.00 $\leq$ log $\theta_{e} \leq$ $-$1.30, where $\theta_{e} \equiv k_{B}T/mc^{2}$, $m$ being the electron mass.  The present paper is organized as follows.  In $\S$2 we will calculate the nonrelativistic electron-electron thermal bremsstrahlung Gaunt factor as a function of the electron temperature $T$ and the photon angular frequency $\omega$, and we will present its accurate analytic fitting formula.  In $\S$3 we will integrate this Gaunt factor over the photon frequency and calculate the frequency-integrated Gaunt factor.  We will also present its accurate analytic fitting formula.  Concluding remarks will be given in $\S$4.

\section{Frequency-Dependent Gaunt Factor}

  In the field of astrophysics it is customary to express the thermal bremsstrahlung emissivity in terms of the Kramers emissivity (Nozawa, Itoh, \& Kohyama~\cite{ref:nozawa}; Itoh et al.~\cite{ref:itoh00}; Itoh et al.~\cite{ref:itoh01}):
\begin{eqnarray}
< W(\omega) >_{K} d \omega & = & \frac{2^{5} \pi e^{6}}{3 h m c^{3}} \, n_{e} n_{j} Z_{j}^{2} \, \left( \frac{2 \pi k_{B} T}{3 m} \right)^{1/2} \, e^{-u} \, \frac{\hbar}{k_{B}T} \, d \omega \, \nonumber  \\
& = & 1.426 \times 10^{-27} \, [ n_{e} (\mbox{cm$^{-3}$}) ] [n_{j} (\mbox{cm$^{-3}$}) ] Z_{j}^{2} [ T(\mbox{K}) ]^{1/2} \, \nonumber  \\
& & \times e^{-u} \, du  \, \, \, \, \, \, \mbox{ergs \, s$^{-1}$ cm$^{-3}$}  \,  ,  \\
u & \equiv & \frac{\hbar \omega}{k_{B}T}  \, ,
\end{eqnarray}
where $\omega$ is the angular frequency of the emitted photon, $T$ is the electron temperature (in kelvins), $n_{e}$ is the number density of the electrons (in cm$^{-3}$), and $n_{j}$ is the number density of the ions with the charge $Z_{j}$ (in cm$^{-3}$).  Maxon \& Corman~\cite{ref:maxon67} and Maxon~\cite{ref:maxon72} have presented the results of the calculation of the nonrelativistic electron-electron thermal bremsstrahlung.  Haug~\cite{ref:haug} has confirmed that their nonrelativistic results agree with the results of the relativistic calculation within about 1\% accuracy at $T \sim 1.5 \times 10^{8}$K.  Since the electron-electron thermal bremsstrahlung makes a percent-order contribution to the total thermal bremsstrahlung at $T \sim 10^{8}$K, the nonrelativistic results of the electron-electron thermal bremsstrahlung will be sufficiently accurate for the analysis of the X-ray observational data of the galaxy clusters.

  Rewriting the results of Maxon \& Corman~\cite{ref:maxon67} and Maxon~\cite{ref:maxon72}, we define the electron-electron thermal bremsstrahlung emissivity by
\begin{eqnarray}
< W(\omega) >_{ee} d \omega & \equiv & g_{ee}(\theta_{e}, u) \, < W(\omega) >_{K} \, d \omega \, ,  \\
g _{ee}(\theta_{e}, u) & = & \frac{n_{e}}{n_{j} Z_{j}^{2}} \, \theta_{e} \, J(\theta_{e}, u) \, , \\
\theta_{e} & \equiv & \frac{k_{B}T}{mc^{2}} \, = \, \frac{T [\mbox{K}]}{5.9299 \times 10^{9}} \, , \\
J(\theta_{e}, u) & = & \frac{ \sqrt{3}}{10 \sqrt{2} \pi} \, u^{2} e^{u} \, I(\theta_{e}, u) \, , \\
I(\theta_{e}, u)  & = &  \int_{0}^{1} \, \frac{dy}{ \sqrt{1-y}}  \, \frac{ \mbox{exp} \left( \pi \alpha \sqrt{ \frac{y}{ \theta_{e} u}} \right) \, - \, 1}{ \mbox{exp} \left( \pi \alpha \sqrt{ \frac{y}{ \theta_{e} u (1-y)}} \right) \, - \, 1} \, A(u, y) \, , \\
A(u, y) & = & \frac{e^{-u/y}}{y^{3}} \, \left[ \left\{ 17 - \frac{3 y^{2}}{(2-y)^{2}} \right\} \sqrt{1-y} \, \right.  \nonumber  \\
&  &  \left. + \left\{ \frac{12 (2-y)^{4} - 7(2-y)^{2} y^{2} - 3y^{4}}{(2-y)^{3}} \right\} \mbox{ln} \left( \frac{1}{\sqrt{y}} + \sqrt{ \frac{1}{y} -1} \right) \right] \, ,  \\
\alpha & = & \frac{1}{137.036}  \, .
\end{eqnarray}

In deriving equation (7), Maxon \& Corman~\cite{ref:maxon67} have taken into account the distortion of the wave function by the Coulomb interaction with the introduction of the Elwert~\cite{ref:elwert} factor.  Nozawa, Itoh, \& Kohyama~\cite{ref:nozawa} have numerically shown that the accuracy of the Elwert approximation is about 0.1\% at $T \sim 10^{8}$K.  Therefore, the Elwert approximation will give sufficiently accurate results for the present purpose.  Equations (3), (4) give the definition of the ``electron-electron thermal bremsstrahlung Gaunt factor" $g_{ee}(\theta_{e},u)$.  We present the results of the calculation of $J(\theta_{e},u)$ in Figures 1, 2.

  We give an analytic fitting formula for $J(\theta_{e},u)$ as follows.  The range of the fitting is $-4.00 \leq$ log $\theta_{e}$ $\leq -1.30$, $-4.00 \leq$ log $u$ $\leq 1.00$.  We express the fitting formula by
\begin{eqnarray}
J(\theta_{e}, u) & = & \sum_{i,k=0}^{10} \, a_{i \, k} \Theta_{e}^{i} \, U^{k} \, , \\
\Theta_{e} & \equiv & \frac{1}{1.35} \, \left( \, \mbox{log} \, \theta_{e} + 2.65 \, \right) \,  , \\
U & \equiv & \frac{1}{2.50} \, \left( \, \mbox{log} \, u + 1.50 \, \right) \, .
\end{eqnarray}
The coefficients $a_{i \, k}$ are presented in Table I.  The accuracy of the fitting is generally better than 0.1\%.

\renewcommand{\baselinestretch}{1.1}
\begin{table}
\caption[]{Coefficients $a_{ik}$}
\begin{tabular}{crrrrrr} \hline

  & {\it k} = 0 \, \, \, & {\it k} = 1 \, \, \, & {\it k} = 2  \, \, \, & {\it k} = 3  \, \, \, & {\it k} = 4  \, \, \, & 
{\it k} = 5  \, \, \, \\ 

 {\it i} &   &   &   &   &   &  \\ \hline

 0  &       3.15847E+0  &    $-$2.52430E+0  &     4.04877E$-$1  &     6.13466E$-$1  &     6.28867E$-$1  &       3.29441E$-$1  \\

 1  &     2.46819E$-$2  &     1.03924E$-$1  &     1.98935E$-$1  &     2.18843E$-$1  &     1.20482E$-$1  &  $-$4.82390E$-$2    \\

 2  &  $-$2.11118E$-$2  &  $-$8.53821E$-$2  &  $-$1.52444E$-$1  &  $-$1.45660E$-$1  &  $-$4.63705E$-$2  &     8.16592E$-$2    \\

 3  &     1.24009E$-$2  &     4.73623E$-$2  &     7.51656E$-$2  &     5.07201E$-$2  &  $-$2.25247E$-$2  &  $-$8.17151E$-$2    \\

 4  &  $-$5.41633E$-$3  &  $-$1.91406E$-$2  &  $-$2.58034E$-$2  &  $-$2.23048E$-$3  &     5.07325E$-$2  &     5.94414E$-$2    \\

 5  &     1.70070E$-$3  &     5.39773E$-$3  &     4.13361E$-$3  &  $-$1.14273E$-$2  &  $-$3.23280E$-$2  &  $-$2.19399E$-$2    \\

 6  &  $-$3.05111E$-$4  &  $-$7.26681E$-$4  &     4.67015E$-$3  &     1.24789E$-$2  &  $-$1.16976E$-$2  &  $-$1.13488E$-$2    \\

 7  &  $-$1.21721E$-$4  &  $-$7.47266E$-$4  &  $-$2.20675E$-$3  &  $-$2.74351E$-$3  &  $-$1.00402E$-$3  &  $-$2.38863E$-$3    \\

 8  &     1.77611E$-$4  &     8.73517E$-$4  &  $-$2.67582E$-$3  &  $-$4.57871E$-$3  &     2.96622E$-$2  &     1.89850E$-$2    \\

 9  &  $-$2.05480E$-$5  &  $-$6.92284E$-$5  &     2.95254E$-$5  &  $-$1.70374E$-$4  &  $-$5.43191E$-$4  &     2.50978E$-$3    \\

10  &  $-$3.58754E$-$5  &  $-$1.80305E$-$4  &     1.40751E$-$3  &     2.06757E$-$3  &  $-$1.23098E$-$2  &  $-$8.81767E$-$3    \\  \hline

\end{tabular}

\smallskip

\begin{tabular}{crrrrr} \hline

  & {\it k} = 6 \, \, \, & {\it k} = 7 \, \, \, & {\it k} = 8  \, \, \, & {\it k} = 9  \, \, \, & {\it k} = 10   \, \, \, \\ 

 {\it i} &   &   &   &    &  \\ \hline

 0  &  $-$1.71486E$-$01  &  $-$3.68685E$-$01  &  $-$7.59200E$-$02  &     1.60187E$-$01  &     8.37729E$-$02  \\
 1  &  $-$1.20811E$-$01  &  $-$4.46133E$-$04  &     8.88749E$-$02  &     2.50320E$-$02  &  $-$1.28900E$-$02  \\
 2  &     9.87296E$-$02  &  $-$3.24743E$-$02  &  $-$8.82637E$-$02  &  $-$7.52221E$-$03  &     1.99419E$-$02  \\
 3  &  $-$4.59297E$-$02  &     5.05096E$-$02  &     5.58818E$-$02  &  $-$9.11885E$-$03  &  $-$1.71348E$-$02  \\
 4  &  $-$2.11247E$-$02  &  $-$5.05387E$-$02  &     9.20453E$-$03  &     1.67321E$-$02  &  $-$3.47663E$-$03  \\
 5  &     1.76310E$-$02  &     2.23352E$-$02  &  $-$4.59817E$-$03  &  $-$8.24286E$-$03  &  $-$3.90032E$-$04  \\
 6  &     6.31446E$-$02  &     1.33830E$-$02  &  $-$8.54735E$-$02  &  $-$6.47349E$-$03  &  3.72266E$-$02  \\
 7  &  $-$2.28987E$-$03  &     7.79323E$-$03  &     7.98332E$-$03  &  $-$3.80435E$-$03  &  $-$4.25035E$-$03  \\
 8  &  $-$8.84093E$-$02  &  $-$2.93629E$-$02  &     1.02966E$-$01  &     1.38957E$-$02  &  $-$4.22093E$-$02  \\
 9  &     4.45570E$-$03  &  $-$2.80083E$-$03  &  $-$5.68093E$-$03  &     1.10618E$-$03  &     2.33625E$-$03  \\
10  &     3.46210E$-$02  &     1.23727E$-$02  &  $-$4.04801E$-$02  &  $-$5.68689E$-$03  &     1.66733E$-$02  \\  \hline

\end{tabular}
\end{table}

\renewcommand{\baselinestretch}{1.5}

\section{Frequency Integrated Gaunt Factor}

  In this section we will integrate the frequency-dependent electron-electron thermal bremsstrahlung emissivity over the whole frequency range.  Thus we obtain
\begin{eqnarray}
W_{ee}(\theta_{e}) & \equiv & \int_{0}^{\infty} < W(u) >_{ee} \, d u  \nonumber  \\
& = & 1.426 \times 10^{-27} \, [ n_{e} (\mbox{cm$^{-3}$})]^{2} [ T(\mbox{K}) ]^{1/2} \, \theta_{e} J(\theta_{e}) \, \, \, \, \, \, \mbox{ergs \, s$^{-1}$ cm$^{-3}$}  \, ,  \\
J(\theta_{e}) & = & \int_{0}^{\infty} e^{-u} J(\theta_{e}, u) \, d u  \nonumber \\
& = & \frac{ \sqrt{3}}{10 \sqrt{2} \pi} \, \int_{0}^{\infty} u^{2} \, I(\theta_{e}, u) \, d u \, .
\end{eqnarray}
We present the results of the calculation of $J(\theta_{e})$ in Figure 3.

  We give an analytic fitting formula for $J(\theta_{e})$ as follows.  The range of the fitting is $-4.00 \leq$ \rm{log} $\theta_{e} \leq -1.30$.  We express the fitting formula by
\begin{eqnarray}
J(\theta_{e}) & = & \sum_{i=0}^{10} \, b_{i} \Theta_{e}^{i} \, , \\
\Theta_{e} & \equiv & \frac{1}{1.35} \, \left( \, \mbox{log} \, \theta_{e} + 2.65 \, \right) \, .
\end{eqnarray}
The coefficients $b_{i}$ are presented in Table II.  The accuracy of the fitting is generally better than 0.1\%.

\renewcommand{\baselinestretch}{1.1}
\begin{table}
\caption[]{Coefficients $b_{i}$}
\begin{tabular}{cr} \hline

  {\it i}  &   $b_{i}$  \, \, \, \, \,   \\  \hline
    0  &      2.21564E+0     \\
    1  &      1.83879E$-$1   \\
    2  &   $-$1.33575E$-$1   \\
    3  &      5.89871E$-$2   \\
    4  &   $-$1.45904E$-$2   \\
    5  &   $-$7.10244E$-$4   \\
    6  &      2.80940E$-$3   \\
    7  &   $-$1.70485E$-$3   \\
    8  &      5.26075E$-$4   \\
    9  &      9.94159E$-$5   \\
   10  &   $-$1.06851E$-$4   \\   \hline

\end{tabular}
\end{table}

\renewcommand{\baselinestretch}{1.5}

\section{Concluding Remarks}

  We have calculated the nonrelativistic electron-electron thermal bremsstrahlung Gaunt factors and have presented their accurate analytic fitting formulae.  The analytic fitting formulae generally have accuracy better than 0.1\%.

  By combining the present electron-electron thermal bremsstrahlung Gaunt factors with the accurate electron-ion thermal bremsstrahlung Gaunt factors which the present authors have calculated in their recent papers, X-ray astronomers will be provided with the precision thermal bremsstrahlung Gaunt factors with the general accuracy better than 1\% for the temperature range 6.0 $\leq$ log $T$[K] $\leq$ 8.5.  These precision thermal bremsstrahlung Gaunt factors will be extremely useful for the analysis of the precision observational data taken by the {\it Chandra X-Ray Observatory} and {\it XMM-Newton}, the X-ray data of high-temperature galaxy clusters in particular. 

\acknowledgments

  We thank Professor Y. Oyanagi for allowing us to use the least square fitting program SALS.  This work is financially supported in part by the Grant-in-Aid of Japanese Ministry of Education, Science, Sports, and Culture under the contract \#13640245.

\newpage

\noindent
{\bf \large{Figure Captions}}

\noindent
Fig.1: The function $J(\theta_{e},u)$ as a function of $u$ for $\theta_{e}$ = 0.01, 0.03.

\noindent
Fig.2: The function $J(\theta_{e},u)$ as a function of $\theta_{e}$ for log$_{10}u$ = $-$4.0, $-$3.0, $-$2.0, $-$1.0, 0.0, 1.0.

\noindent
Fig.3: The function $J(\theta_{e})$ as a function of $\theta_{e}$.


\begin{thebibliography}{0}

\bibitem{ref:allen} \BY{Allen~S.~W., Ettori~S. \atque Fabian~A.~C.} {\it MNRAS}, in press (2001)

\bibitem{ref:schmidt} \BY{Schmidt~R.~W., Allen~S.~W. \atque Fabian~A.~C.} {\it MNRAS}, in press (2001)

\bibitem{ref:nozawa} \BY{Nozawa~S., Itoh~N. \atque Kohyama~Y.} \IN{ApJ}{507}{1998}{530}

\bibitem{ref:itoh00} \BY{Itoh~N., Sakamoto~T., Kusano~S., Nozawa~S. \atque Kohyama~Y.} \IN{ApJS}{128}{2000}{125}

\bibitem{ref:itoh01} \BY{Itoh~N., Sakamoto~T., Kusano~S., Kawana~Y. \atque Nozawa~S.} A\&A, in press (2002)

\bibitem{ref:itoh85} \BY{Itoh~N., Nakagawa~M. \atque Kohyama~Y.} \IN{ApJ}{294}{1985}{17}

\bibitem{ref:nakagawa} \BY{Nakagawa~M., Kohyama~Y. \atque Itoh~N.} \IN{ApJS}{63}{1987}{661}

\bibitem{ref:itoh90} \BY{Itoh~N., Kojo~K. \atque Nakagawa~M.} \IN{ApJS}{502}{1990}{7}

\bibitem{ref:itoh91} \BY{Itoh~N., Kuwashima~F., Ichihashi~K. \atque Mutoh~H.} \IN{ApJ}{382}{1991}{636}

\bibitem{ref:itoh97} \BY{Itoh~N., Kojo~K., Nakagawa~M., et al.}, in AAS CD-ROM Series, Astrophysics on Discs, Vol.9 (1997, Washington, D.C.: AAS)

\bibitem{ref:maxon67} \BY{Maxon~M.~S. \atque Corman~E.~G.} \IN{Phys. Rev.}{163}{1967}{156}

\bibitem{ref:maxon72} \BY{Maxon~M.~S.} \IN{Phys. Rev.}{A5}{1972}{1630}

\bibitem{ref:haug} \BY{Haug~E.} \IN{Z. Naturforsch.}{30a}{1975}{1546}

\bibitem{ref:svensson} \BY{Svensson~R.} \IN{ApJ}{258}{1982}{335}

\bibitem{ref:dermer} \BY{Dermer~C.~D.} \IN{ApJ}{307}{1986}{47}

\bibitem{ref:elwert} \BY{Elwert~G.} \IN{Ann. Phys.}{34}{1939}{178}

\end{thebibliography}
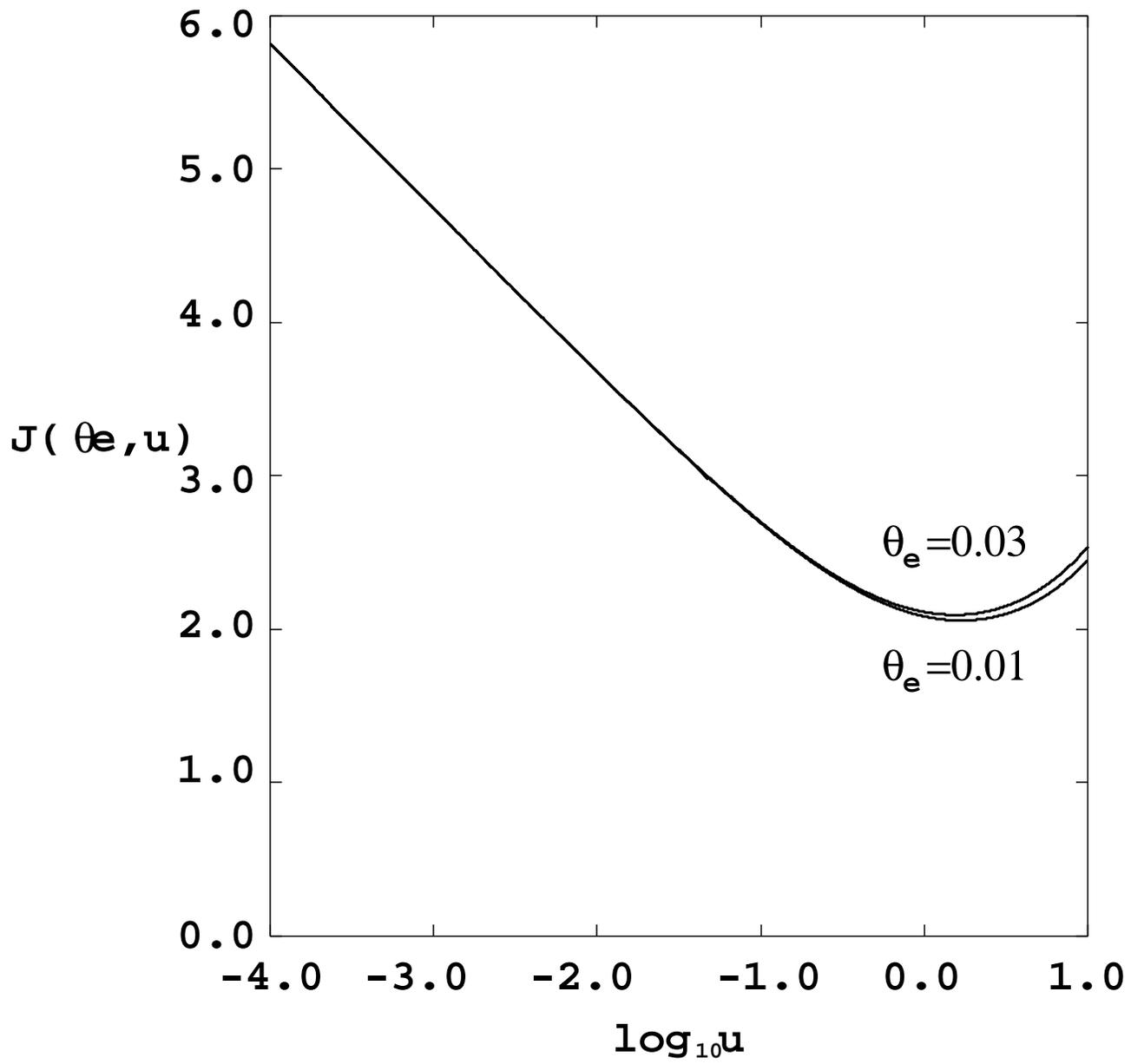
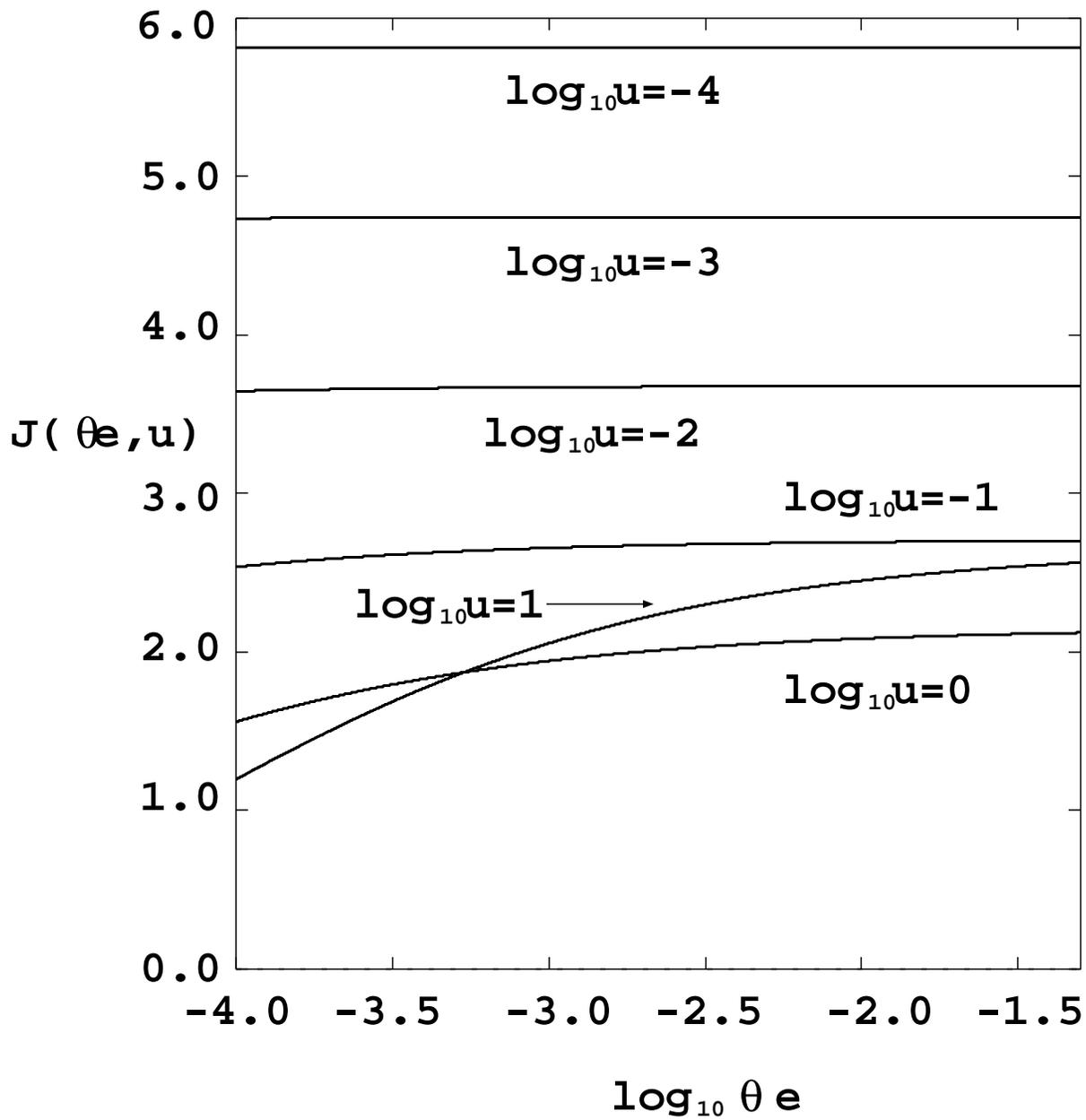
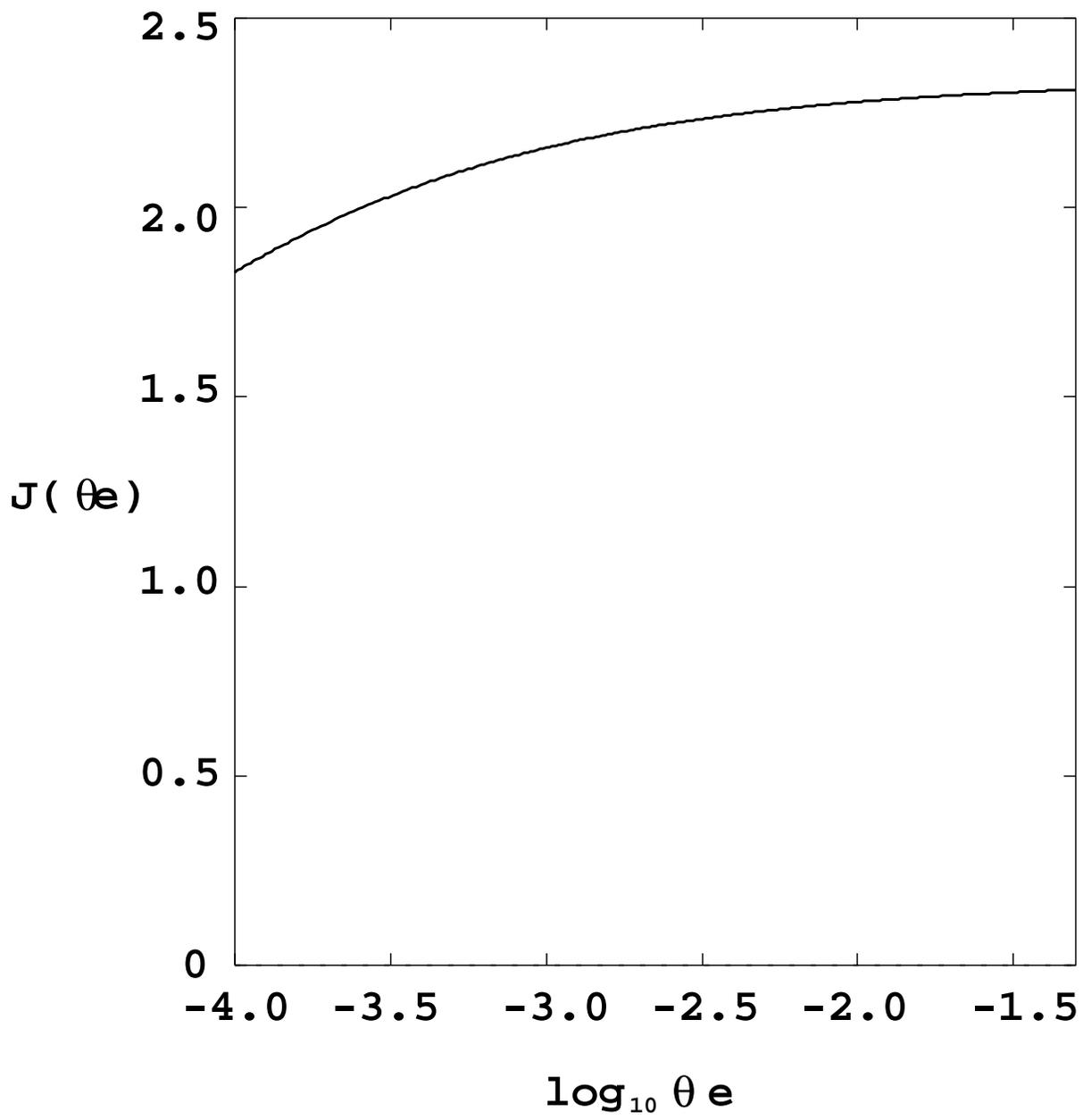
\end{document}